\documentclass[aps,pra,twocolumn,showpacs,superscriptaddress,amsmath,amssymb]{revtex4}
\usepackage{graphicx,epsfig}
\begin{document}


\title{On the Salecker-Wigner-Peres clock and double barrier tunneling}



\author{Marcos Cal\c cada}
\email[]{mcalcada@uepg.br}

\author{Jos\'{e} T. Lunardi}
\email[]{jttlunardi@uepg.br}
\affiliation{Grupo de F\'{\i}sica Te\'orica e Modelagem
Matem\'atica, Departamento de Matem\'atica e Estat\'{\i}stica,
Universidade Estadual de Ponta Grossa. Av. General Carlos
Cavalcanti, 4748\\ 84030-000, Ponta Grossa, PR, Brazil}

\author{Luiz A. Manzoni}
\email[]{manzoni@cord.edu}
\affiliation{Department of Physics, Concordia
College, 901 8th St. S., Moorhead, MN 56562, USA}


\date{\today}

\begin{abstract}
In this work we revisit the Salecker-Wigner-Peres clock formalism and show that it can be directly applied to the phenomenon of tunneling. Then we apply this formalism to the determination of the tunneling time of a non relativistic wavepacket, sharply concentrated around a tunneling energy, incident on a symmetric double barrier potential. In order to deepen the discussion about the generalized Hartmann effect, we consider the case in which the clock runs only when the particle can be found inside the region \emph{between} the barriers and show that, whenever the probability to find the particle in this region is non negligible, the corresponding time (which in this case turns out to be a dwell time) increases with the barrier spacing.
\end{abstract}


\maketitle


\section{Introduction}

An unambiguous definition of a tunneling time is an important problem in
quantum mechanics, due to both its applications and its relevance to the foundations
of the theory. It is, however, a problem which has eluded
physicists since the beginnings of the quantum theory. Many attempts have been made to define such a time scale (see
\cite{HSt89, LMa94, Win06-2, Win06-1} for reviews). However, most of
these definitions (phase time, dwell time, Larmor time, etc.), while
being valid to describe some specific characteristic of the tunneling
process, also present difficulties if one tries to interpret them
as traversal times in general.

Perhaps the most striking of the above mentioned difficulties concerns the issue of superluminality,
a direct consequence of the Hartmann effect \cite{Har62}, which asserts that the phase time saturates for opaque barriers. More recently, the Hartmann effect has been considered for the double barrier potential, and it
was verified that in the opaque limit the phase time does not depend
on the spacing between the barriers either, a phenomenon referred to as the \textit{generalized Hartman effect} \cite{ORS02} (also see \cite{LLB02, Esp03, DLR05}). Although there are no real paradoxes associated with these phenomena, since it is well known that the group velocity cannot be associated to the signal velocity in such a situation \cite{SB}, the subject has originated intense debate in the literature (see, for example, \cite{Mil05} and the references there cited). Independently of these controversial interpretations, it is in fact a counterintuitive result that in the opaque limit the phase time does not depend on the spacing between the barriers, because one could expect group velocity to have the usual physical meaning in that \emph{free} region. Subsequent investigations, both in the non-relativistic \cite{Win05} and in the relativistic \cite{LMa07} cases, indicate that this lack of dependence may be an artifact of the opaque limit, but the subject still deserves further investigation.

A fruitful avenue of investigation on tunneling times considers the use of quantum clocks. A quantum clock is a secondary dynamical system weakly coupled to the system of interest and having a degree of freedom evolving uniformly in time. One of the most prominent works along this line leads to the Larmor time \cite{Larmor, FHa88}, but other clocks are possible (see for example, \cite{LMa94} and references there cited). Here we are particularly interested in the clock formalism introduced by Salecker and Wigner \cite{SWi58} and later revisited by Peres \cite{Per80}, who used it to investigate, among other problems, the time-of-flight for a non-relativistic particle (see also \cite{AMM03}). The extension for a relativistic particle was later done by Davies \cite{Dav86}.

In \cite{Per80} Peres also introduced a ``time operator" (\emph{not} canonically conjugate to the clock's Hamiltonian), whose expectation values do not lead to sensible results for the tunneling time in the presence of a localized potential, as was later shown by Leavens \emph{et al} \cite{Lea93, LMc94}. To overcome such a difficulty, these authors proposed a modification of the original Salecker-Wigner-Peres (SWP) formalism by the introduction of a calibration procedure. However, in his treatment of the time-of-flight problem Peres \cite{Per80} did not use directly such an operator, but defined the time given by the clock ($t_c$) as the derivative of the phase shift of the wavefunction with respect to the perturbation potential. In this work we demonstrate that Peres' original approach, contrary to what is usually stated, can be directly applied to the tunneling time problem, without the need for calibration. In section \ref{swp} we present a brief review of the SWP formalism and clarify some important issues related to it. In particular, we present a simple proof of the general result that $t_c$ (averaged over the scattering channels) is exactly equal to the well known dwell time (a result obtained by Leavens through the expectation values of the Peres' ``time operator'' only \emph{after} calibration). In section \ref{db} we apply the SWP formalism to the tunneling through a symmetric double barrier and analyze the dependence of $t_c$ with the spacing between the barriers. It must be noticed that although in this case the (transmitted or reflected) time resulting from the SWP clock is exactly the dwell time, this formalism proves to be operationally better suited to address the question of independence or not of the tunneling time with respect to the barrier spacing in the limit of opaque barriers, providing a simpler procedure for the direct calculation of the time spent by the wavepacket only \emph{between} the two barriers. Therefore, such approach allows us to deepen the discussions about the generalized Hartmann effect. In Section \ref{concl} we discuss the results and their interpretation. In the Appendix we list the explicit expressions for some terms appearing in the expressions for the times obtained in section \ref{db}.


\section{The SWP clock and the tunneling time problem}
\label{swp}

The \emph{free} SWP clock consists of a quantum rotor, which for a Hilbert space of dimension $N$ has Hamiltonian given by \cite{Per80}
\begin{equation}
\label{cham}
H_c=\omega J\, ,
\end{equation}
with $J\!=\!-i\hbar \frac{\partial}{\partial \theta}$, $\omega\!=\!\frac{2\pi}{N\tau}$ and $\tau$ is the clock's resolution.
The energy eigenstates are $u_m\!=\!\frac{\mathrm{e}^{i m \theta}}{\sqrt{2\pi}}$, $\theta\! \in \![0,2\pi )$, with eigenvalues $\mathcal{V}_m\!=\!m \hbar\omega $ \,($m\!=\!-j,...,j$).
Another convenient orthonormal basis for the clock's Hilbert space is \cite{Per80}
\begin{equation}
\label{v}
v_k(\theta)=\!\frac{1}{\sqrt{2\pi N}}\,\frac{\sin \left[\frac{N}{2}\left(\theta-\omega k\tau\right)\right]}{\sin \left[\frac{1}{2}\left(\theta-\omega k\tau\right)\right]}, \hspace{0.5cm} \theta \in [0,2\pi )
\end{equation}
with $k\!=\!0,...,N\!\!-\!\!1 $. These are the eigenstates of the Hermitian operator
\begin{equation}
\label{T}
T= \sum_{k=0}^{N-1} k\tau {\cal P}_{k}\, , \qquad \mathrm{with}\quad {\cal P}_{k}v_l=\delta_{kl}v_l,
\end{equation}
with eigenvalues $k\tau$. The above operator plays the role of a ``time" operator, despite not being canonically conjugate to the Hamiltonian. The motivation for this identification is that, for large $N$, the
wavefunctions $v_k$ are sharply peaked around $\theta=\omega k\tau$, with a width $ \frac{2\pi}{N}$, and their time evolution is given by
\begin{equation}
\label{vt}
\exp\left(-\frac{i}{\hbar} H_c t\right) v_k(\theta)=v_k(\theta-\omega t).
\end{equation}
Thus, $v_k(\theta)$ evolves \emph{rigidly} and \emph{uniformly} within the interval $[0,2\pi)$. In particular, for large $N$ the peaks translate from $\omega k \tau$ to $\omega (k\tau+t)$ (mod $2\pi$).

It must be noticed that it is only for times $t=n\tau$, with $n$ an integer, that
$$
v_k(\theta-\omega n\tau)=v_{(k+n)\,(\mathrm{mod}\, N)}(\theta),
$$
such that the whole set of eigenfunctions $v_k(\theta)$ can be obtained from any of them (say $v_0$) through a sequence of (discrete) time translations. This fact is at the origin of the discrepancies found by Leavens \cite{Lea93} between the \emph{intrinsic} time $t$  in (\ref{vt}) and the expectation value of (\ref{T}) whenever $t$ is not an integer multiple of $\tau$. To overcome such a difficulty, Leavens introduced a \emph{calibration procedure}, later revised by Leavens and McKinnon \cite{LMc94}, designed in such a way that after calibration the clock times (given as an average over an ensemble of freely running clocks) coincided with the intrinsic ones.

On the other hand, the above properties of the wavefunctions $v_k$ under time evolution allowed Peres to consider them as
the proper clock's hand, with the clock's ``reading" given by the angle $\omega t_c$ (the translation of the wavefunction's peak).
In applying this approach to the one-dimensional scattering of a particle of mass $\mu$ by a localized potential $V(z)$ confined within the region $0\!<\!z\!<\!L$,
the clock-system coupling can be designed to measure the time the wavepacket spends within an \emph{arbitrary} region $z_1\!<\!z\!<\!z_2$.
In this case the Hamiltonian for the  coupled system is given by \cite{Per80}
\begin{equation}
\label{csham}
H=H_s+P(z)H_c,
\end{equation}
where $H_s=\frac{p^2}{2\mu}+V(z)$, $H_c$ is the clock Hamiltonian (\ref{cham}) and $P(z)$ is a projection operator into the interval $(z_1,z_2)$.
Let us consider a particle in a stationary state of energy $E$ incident from the left
(the results remain valid for a wavepacket strongly concentrated around $E$).
For the initial (free) clock state we choose, following Peres, $v_0(\theta)$.
Then, assuming that the highest eigenvalue of the clock, $\mathcal{V}_m$, is negligible when compared to all the relevant
energy scales in the problem, the \emph{final} (asymptotic) state of the whole system is given by \cite{Per80, Lea93}
\begin{equation}
\Psi(z,\theta)\!=\!
\left\{
\begin{array}{l} T \; \mathrm{e}^{i k z} \;v_0\left(\theta -\omega \;t^T_c\right),\quad z\geq\mathrm{max}\left\{z_2,L\right\} \\
\\
R \;\mathrm{e}^{-i k z}\; v_0\left(\theta -\omega\; t^R_c\right),\quad z\leq\mathrm{min}\left\{z_1,0\right\},
\end{array}
\right.\nonumber
\end{equation}
where $T$ and $R$ are the transmission and reflection coefficients (which depend on the energy $E$) for the system in the absence
of the clock. From the above expressions and (\ref{vt}), one identifies the \emph{Peres' transmission}
and \emph{reflection times}, respectively, by
\begin{equation}
\label{tp}
t^T_c(E)\!=\!\left.-\hbar\frac{\partial \phi_T^{(m)}}{\partial \mathcal{V}_m}\right|_{\mathcal{V}_m=0}\quad \mathrm{and} \quad
t^R_c(E)\!=\!\left.-\hbar\frac{\partial \phi_R^{(m)}}{\partial \mathcal{V}_m}\right|_{\mathcal{V}_m=0}\!,
\end{equation}
where $\phi_T^{(m)}$($\phi_R^{(m)}$) is the phase delay of transmission(reflection) in the presence of the clock, and the superscript $(m)$ indicates the $m$-th clock's eigenstate. In deriving the above result it is assumed that in the vanishingly weak coupling limit  $T^{(m)}\approx \left|T\right|\mathrm{e}^{i \phi_T^{(m)}}$ and $R^{(m)}\approx \left|R\right|\mathrm{e}^{i \phi_R^{(m)}}$ \cite{Per80}.

A relation between the dwell time  and (\ref{tp}) can be obtained by following steps similar to those Winful used to derive a relation between the dwell time and the phase times \cite{Win03}
(see also \cite{Win04}). In order to do this,  one must realize that for a stationary incident particle and when the clock is in its $m$-th stationary state the problem
is reduced to the solution of the time-independent Schr\"{o}dinger equation with a localized potential
$V^{(m)}(z)\! =\! V(z) + \mathcal{V}_m \mathcal{P}(z)$ \cite{Per80,Lea93}, whose solution outside the potential
region is given by
\begin{equation}
\psi^{(m)}(z)=\left\{
\begin{array}{ll}
T^{(m)}\mathrm{e}^{i k z}, & z\geq\mathrm{max}\left\{z_2,L\right\} \\
& \\ \label{pm}
\mathrm{e}^{i k z}+R^{(m)} \mathrm{e}^{-ik z},& z\leq\mathrm{min}\left\{z_1,0\right\}\; ,
\end{array}
\right.
\end{equation}
where $k=\frac{1}{\hbar}\sqrt{2\mu E}$. Considering the Schr\"odinger equation with the potential $V^{(m)}(z)$ and its complex conjugate we obtain, after
taking the vanishingly weak coupling limit $\mathcal{V}_m\!\to\!0$,
\begin{eqnarray}
\mathcal{P}(z)\,\psi^*\psi&\!=\!&-\frac{\hbar^2}{2\mu}\frac{\partial}{\partial z}
\left[
\left(\frac{\partial \psi^*}{\partial z}\right)\left( \frac{\partial \psi^{(m)}}{\partial \mathcal{V}_m}
\right)_{\mathcal{V}_m=0}\right.\nonumber\\
&&\left.-\psi^* \left(\frac{\partial^2 \psi^{(m)}}{\partial \mathcal{V}_m \partial z}
\right)_{\mathcal{V}_m=0} \right],\label{integ}
\end{eqnarray}
where $\psi$ denotes the wavefunction after the limit $\mathcal{V}_m\!\to\!0$. Integrating the above expression over the region $(z_1,z_2)$ we obtain $\left.\frac{-2\mu}{\hbar ^2}\int_{z_1}^{z_2}\!\! dz |\psi|^2=[...]_{z_2}-[...]_{z_1}\right.$, where $[...]$
corresponds to the term into brackets in the above expression, which is constant for all $z\geq z_2$ and for all
$z\leq z_1$, because $\mathcal{P}(z)=0$ in those regions. Taking advantage of this fact,
we can use any value of $z>z_2$ to compute the bracket $[...]_{z_2}$. For convenience we choose $z$ into
the region corresponding to the transmitted wave. In the same way, we can choose $z$ into the
incident/reflection region to compute the bracket $[...]_{z_1}$. This procedure, together with (\ref{pm}),
yields
$$
\frac{\mu}{i\hbar^2 k}\int_{z_1}^{z_2}\!\! dz |\psi|^2\! = \! T^*\left(\frac{\partial T^{(m)}}{\partial \mathcal{V}_m}\right)_{\!\mathcal{V}_m=0}\!\! \!\!\!\!\! + R^*\left(\frac{\partial R^{(m)}}{\partial \mathcal{V}_m}\right)_{\!\mathcal{V}_m=0}\, , \!\!\!\!
$$
which can be rewritten as
$$
\frac{\mu}{\hbar k}\int_{z_1}^{z_2}\!\!\! dz |\psi|^2 \! = \!
|T|^2\left[-\hbar\frac{\partial \phi_T^{(m)} }{\partial \mathcal{V}_m}\right]_{\!\mathcal{V}_m=0}\!\! \!\!\!\!\!\!\!\!\!+ |R|^2\left[-\hbar\frac{\partial \phi_R^{(m)} }{\partial \mathcal{V}_m}\right]_{\!\mathcal{V}_m=0}\!\!\!\!\!\! .
$$
Identifying the incident flux $j_{in}=\frac{\hbar k}{\mu}$, the l.h.s. of the above expression is the well known
expression for the dwell time $t_D$ \cite{HSt89,LMa94}. Finally, from (\ref{tp}) we obtain
\begin{equation}
\label{relt}
t_D=|T|^2 t^T_c + |R|^2t^R_c \; .
\end{equation}
Although the above relation was also obtained by Sokolovski \emph{et al.} \cite{SokoBaskin} through the use of Feynman's path integrals, our proof is worth mentioning due to its simplicity.  Analogous results were also obtained in the framework of weak measurements or through the Larmor clock formalism, but involving, in general, complex times \cite{Ste, IannFP, IannWM}. Relation (\ref{relt}), however, involves only real times. Leavens and McKinnon \cite{LMc94} showed that using the ``time operator", eq. (\ref{T}), such relation can only be obtained after applying their calibration procedure. As another important point concerning the above relation we emphasize that, \emph{as long as} the transmission and reflection times are defined by (\ref{tp}), no interference term enter it (the analogous relation involving phase times necessarily requires such a term \cite{Win03}). The validity of such a relation has been much debated in the literature (see, for instance, \cite{HSt89, LMa94, LAe89} for different points of view) and we hope that the above derivation helps to clarify the fact that it all depends on how the transmission and reflection times are defined (see also \cite{Ste, Win06-2} for further discussions).

Finally, we note that when the whole potential $V(z)\!+\!\mathcal{V}_m \mathcal{P}(z)$ is \emph{symmetric} the reflected and the transmitted phases differ only by a constant \cite{FHa88}, which leads to $t_D=t_c^T=t_c^R$  \cite{Win06-1,LMa07}. In such a case it must also be noted that  any of the expressions in (\ref{tp}) constitute an operationally simpler way to calculate the exact dwell time. We shall take advantage of this fact in the next section.


\section{The Double Barrier Tunneling}
\label{db}

Let us now consider a particle having a given energy $E$ (or a wavepacket sharply concentrated around this energy) incident from the left on a symmetric double barrier potential, given by
\begin{equation}
\label{Vdouble}
V(z) = V_0\left\{ \Theta (z) \Theta(a-z) + \Theta (z-d-a) \Theta(d+2a-z) \right\}.
\end{equation}
We will consider only the case $E<V_0$, characterizing a tunneling process. Even though we are chiefly concerned with the case in which the clock runs only if the particle is inside the region separating the two barriers ($a<z<a+d$), it is instructive to first consider the clock running when the particle is anywhere within the potential region $(0,2a\!+\!d)$. The solution of the time-independent Schr\"{o}dinger equation outside the potential region in this case is of the form (\ref{pm}), with
the transmission amplitude given by
\begin{widetext}
\begin{eqnarray}
\nonumber  T^{(m)} &\!=\!& 8ikp_mq_m^2 e^{-i(2a+d)k}\left\{ 2\sin (p_md)(k^2+q_m^2)(p_m^2+q_m^2)
 - \left[ (k-iq_m)^2 \left( 2p_mq_m\cos(p_md) \right. \right. \right. \\
 &&+\!\left.\left.\left. (p_m^2-q_m^2)\sin (p_md) \right)e^{-2q_ma} + (q_m - ik)^2 \left( 2p_mq_m\cos(p_md) - (p_m^2 -  q_m^2)\sin (p_md) \right)e^{2q_ma} \right] \right\}^{-1},
\end{eqnarray}
\end{widetext}
where $q_m\!\!=\!\!\sqrt{2\mu (V_0 + \mathcal{V}_m - E)}/\hbar$ and $p_m\!\!=\!\!\sqrt{2\mu (E-\mathcal{V}_m)}/\hbar$.
The corresponding phase is
\begin{equation}\label{fase}
\phi_T^{(m)} = -(2a+d)k -\tan^{-1}\left(\frac{\beta_m}{\alpha_m}\right),
\end{equation}
with $\alpha_m$ and $\beta_m$ defined as
\begin{eqnarray}
\nonumber \alpha_m &\equiv& 2kq_m\left[ 2p_mq_m\cos (p_md) \cosh (2q_ma) \right.\\
\nonumber          &+& \left. (q_m^2-p_m^2)\sin (p_md)\sinh (2q_ma) \right] ;\\  \label{ab} \\
\nonumber  \beta_m &\equiv& - (k^2+q_m^2)(p_m^2+q_m^2)\sin (p_md) \\
\nonumber        &+& 2p_mq_m (q_m^2-k^2)\cos (p_md) \sinh (2q_ma) \\
\nonumber        &+& (q_m^2-p_m^2)(q_m^2-k^2) \sin (p_md)\cosh (2q_ma).
\end{eqnarray}
From  the symmetry of the total potential (including the term due to the clock) it follows that $t_D=t_c^{R}=t_c^{T}$. So, using (\ref{tp}) we obtain
\begin{equation}\label{twhole}
t_D=t_c^{T}= -\frac{\mu}{\hbar \left( \alpha_0^2+\beta_0^2\right)}\left(\frac{h_1}{k}-\frac{h_2}{q}\right),
\end{equation}
where $\alpha_0$ and $\beta_0$ are obtained from (\ref{ab}) by taking the limit $\mathcal{V}_m \!\rightarrow \! 0$ (which corresponds to  $p_m \rightarrow k$ and $q_m \rightarrow q =  \frac{1}{\hbar}\sqrt{2\mu (V_0 - E)}$). The explicit expressions for $\alpha_0$ and $\beta_0$ as well as for $h_1$ and $h_2$ are given in the Appendix.

Now, let us consider the case in which the SWP clock runs only when the particle can be found inside the region \emph{between}
the two barriers, namely in the interval $(a,a\!+\!d)$.  The (transmitted) Peres' time $t_c^{\rm bet}$ in this case can be  obtained from the above results simply
by taking the limit $q_m \!\rightarrow\! q$ in (\ref{fase}) before taking the derivative in (\ref{tp}), which gives
\begin{equation}\label{tbet}
t_c^{\rm bet} = - \frac{\mu}{\hbar k}\frac{h_1}{\alpha_0^2+\beta_0^2}.
\end{equation}
Using the above expression, we can rewrite (\ref{twhole}) as
\begin{equation}\label{tc2}
t_D = t_c^{\rm bet} + t_c^{\rm bar},
\end{equation}
where
\begin{equation}
\label{tbar}
t_c^{\rm bar} =  \frac{\mu}{\hbar q}\frac{h_2}{\alpha_0^2+\beta_0^2}
\end{equation}
is just the (transmitted) Peres' time obtained by allowing the clock to run only when the particle passes \emph{within any} of the two barriers (which corresponds to take $p_m\to k$ in all the above perturbed expressions, before taking the derivative in (\ref{tp})). From the proof presented in the previous section, together with the symmetry of the total potential, it follows that (\ref{tbet}) and (\ref{tbar}) are the \emph{dwell times} spent in the regions \emph{between} and \emph{within} the two barriers, respectively.

In order to discuss the generalized Hartman effect we specialize our results to the opaque limit $qa \to \infty $, in which the Peres' time (\ref{twhole}) for the whole potential region
 \emph{saturates} to the value
\begin{equation}\label{tcOpaq}
t_D=t_c^T \begin{array}{c} \\ _{qa\to\infty}\end{array}\!\!\!\!\!\!\!\!\!\!\!\!\!\!\! \longrightarrow \;
\frac{\mu}{\hbar q^2}\frac{2kq}{(k^2+q^2)}.
\end{equation}
Then, in the opaque limit the behavior of the Peres' time (dwell time) for the \emph{whole} potential region is analogous to that of the phase time, in which it is independent both of the barriers width $a$ and the barrier spacing $a$, and we obtain a version of the generalized Hartman effect. The above saturated result could also be obtained from the non relativistic limit for the dwell time obtained in \cite{LMa07} (in that reference the dwell time was obtained from a relation among the phase and the dwell times involving interference terms \cite{Win04}). However, methods based on the phase time, which is an asymptotically extrapolated quantity, are not suitable to study the behavior of the time the particle spends \emph{only} inside the region \emph{between} the barriers.

\begin{figure}
\includegraphics[width=9.0cm,height=7.0cm]{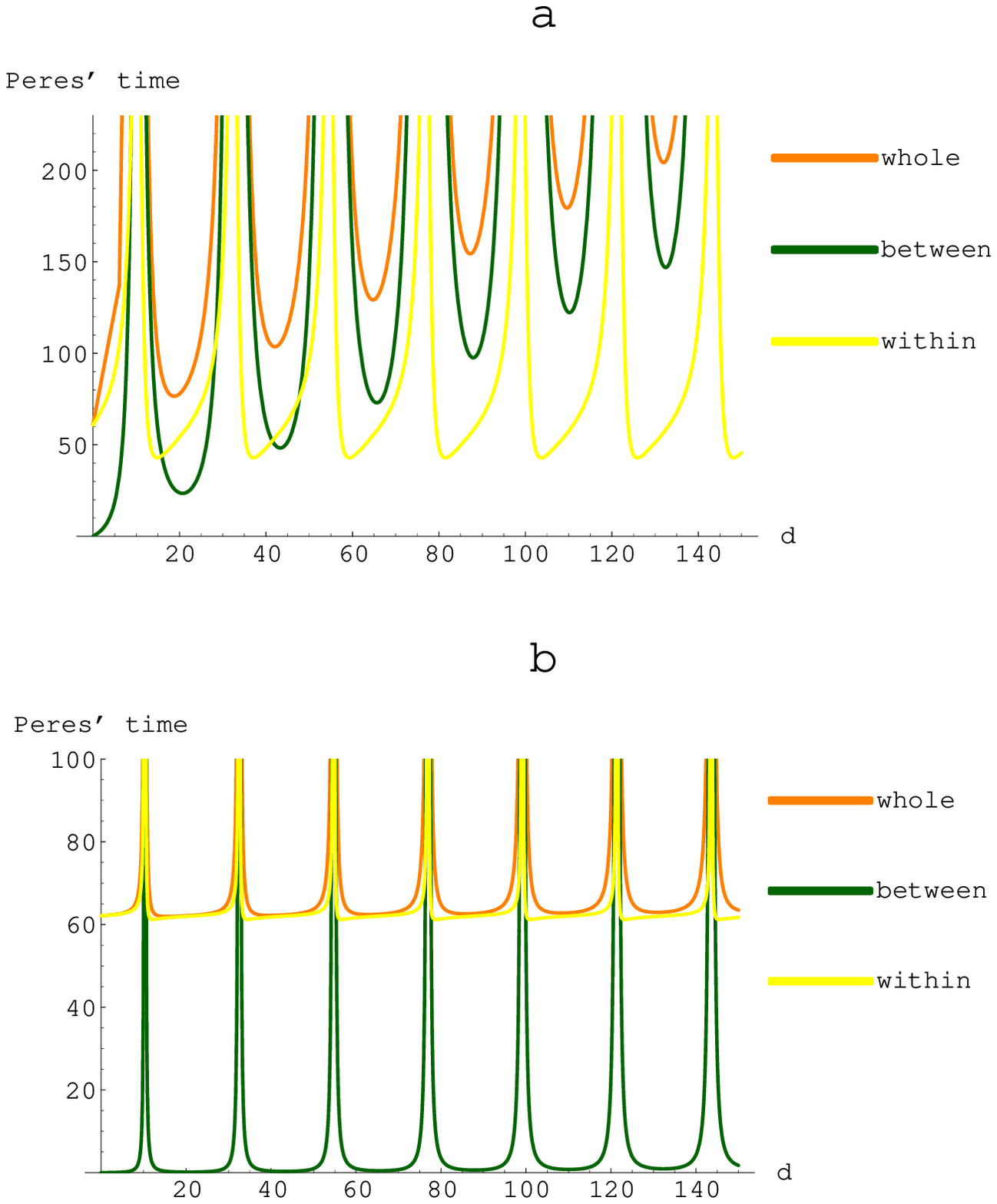}
\caption{(Color online) Peres' clock time dependence with respect to the distance $d$ between the two barriers. In both figures the three clock times appearing in relation (\ref{tc2}) are shown, corresponding to the \emph{whole} potential region $(0,2a+d)$, to the region \emph{between} the barriers $(a,a+d)$, and to the region \emph{within} the two barriers $(0,a)\cup (a+d,2a+d)$. \emph{Natural units} were used ($\hbar=c=1$), so that mass and energies are expressed in units of the particle mass $\mu$, and distances and times are expressed in units of $\mu^{-1}$. In both figures $E=0.01$ and $V_0=0.018$. (a) $a=10$; (b) $a=30$. \label{fig1}}
\end{figure}

In the opaque limit $qa \rightarrow \infty$,  the expression (\ref{tbet}) immediately yields $t_c^{\rm bet} \rightarrow 0$.
A more careful analysis taking into account the leading terms in the asymptotic situation in which $qa$ is large, but finite,
shows that
\begin{widetext}
\begin{equation}
t_c^{\rm bet} \begin{array}{c} \\ _{qa\gg 1}\end{array}\!\!\!\!\!\!\!\!\!\!\!\!\!\!\! \longrightarrow \; \frac{4\mu q^2}{\hbar}\frac{e^{-2qa}}{(k^2+q^2)} \frac{ \left\{2kd (k^2+q^2) + 4kq \sin^2 (kd) + (k^2-q^2) \sin (2kd) \right\}}{\left\{ (k^2-q^2) \sin (kd) -2kq \cos (kd) \right\}^2}.
\end{equation}
\end{widetext}
Apart from terms coming from the multiple reflections and/or interference at the barriers, this expression clearly displays an increasing (almost linear) dependence of $t_c^{\rm bet}$ with respect to the barrier
spacing $d$ , \emph{for any finite} $qa$. Fig. \ref{fig1} shows the behaviors of the three times entering expression (\ref{tc2}),
with increasing $d$. Fig. \ref{fig1}a concerns a large but finite barrier width $a$. We can observe the almost linear increasing of $t^{\rm bet}$ with increasing $d$. In Fig. \ref{fig1}b the barrier width is increased three times and we can already observe the tendency to saturation (except by the peaks) of the three times:
 $t_D$  and $t_c^{\rm bar}$ tends to the saturated value (\ref{tcOpaq}), while  $t_c^{\rm bet}$ tends to saturate to zero. Increasing even more the barrier width makes the peaks vanish. It can also be shown that if the clock runs only within the \emph{first} barrier, the corresponding saturated dwell time in the opaque limit is exactly the same as (\ref{tcOpaq}). Summarizing, in the opaque limit $qa\to \infty$ the transmission amplitude beyond the \emph{first} barrier goes to zero, and its associated phase becomes meaningless. This fact corroborates the view that the generalized Hartman effect, which asserts the independence of the tunneling time on $d$,
 is indeed an artifact of the opaque limit (see also \cite{Win05,LMa07}).

\section{Concluding Remarks}
\label{concl}

In this work we revisited the SWP clock formalism, which (despite of being based on a thought experiment) is useful to understand the fundamental concepts involved in the definition of a time scale for tunneling processes. We showed that in Peres' version it can be directly applied to the problem of tunneling times. We demonstrated that departures from the Peres' approach are not necessary, even in the case of localized potentials, if one focuses, as Peres did, on  the time evolution of the eigenfunctions of his ``time operator", instead of focusing on its expectation values, as did Leavens \emph{et al} \cite{LMc94}.

Using the Peres' approach, and through a simple extension of a proof originally designed by Winful in the context of phase times \cite{Win03,Win04}, we have shown that the Peres' times (\ref{tp}), when weighted by the transmission and reflection probabilities, averages exactly to the dwell time (incidentally, for symmetric potentials any of the two expressions (\ref{tp}) provides an operationally simpler way to calculate the dwell time). On the other hand, we did not address questions regarding the SWP clock's resolution in the presence of a localized potential (see \cite{Per80, Lea93}) because this issue can be addressed in the framework of the weak measurement theory \cite{weak}, as suggested in \cite{Dav04, Dav05}. Besides that, in this paper each of the time readings associated to the clock is equivalent to a dwell time, which  is a well established time scale \cite{HSt89, LMa94, Win06-2, Win06-1}.

We then applied the SWP formalism to the symmetric double barrier potential, aiming to analyze the so-called generalized Hartman effect. We calculated explicitly the dwell time and verified that in the opaque limit it does not depend on the barrier separation, confirming the emergence of the generalized Hartmann effect also in this case (see also \cite{Win05}). However, we added a new insight into this debated question by allowing the clock to run only inside the region between the barriers, and taking into consideration the leading terms when the barrier width is large, but finite, we unambiguously showed that the dwell time increases ``almost linearly" with the barrier spacing $d$ (apart from terms arising from the multiple reflections/interference inside this region). The fact that such a behavior is modulated by an exponential decay $\exp (-2qa)$ can be understood by noticing that the dwell time is an average over the probability of finding the particle in the interest region (and since this probability decays exponentially with $qa$, so will $t_D$). Therefore, whenever the probability to find the particle in the region between the barriers is non negligible, the corresponding dwell time depends on the barrier spacing.

All the above considerations reinforce the conclusions arrived in \cite{Win05}, in the context of a Fabry-P\'erot cavity, and in \cite{LMa07}, for the relativistic tunneling through double barriers, that the generalized Hartman effect is just a mathematical artifact of the opaque limit: although in that limit the transmission phase is well defined  and finite, it is meaningless since $T$ itself goes to zero. Therefore, any time scale defined in terms of such phase (such as phase times and, as seen in section II, dwell times) also becomes meaningless in this limit: it corresponds to the trivial fact that the particle does not penetrate past the first barrier, and therefore it makes no sense to associate any time duration to its passage (or dwelling) in the region between the two barriers.

\appendix*
\section{}

In this Appendix we present the explicit expressions for $\alpha_0$, $\beta_0$, $h_1$, and $h_2$, which appear in the expressions
for the Peres' clock times (\ref{twhole}), (\ref{tbet}), and (\ref{tbar}):
\begin{eqnarray*}
\nonumber
\alpha_0 \!\equiv\!\!&& 2kq\!\left[ 2kq\cos (kd) \cosh (2qa) \right.\\
&&\left.+\! (q^2\!-\!k^2)\sin (kd)\sinh (2qa) \right],
\end{eqnarray*}
\begin{eqnarray*}
\beta_0 \!\equiv\!&& - (k^2\!+\!q^2)^2\sin (kd)\! +\! 2kq (q^2\!-\!k^2)\cos (kd) \sinh (2qa) \\
                 &&+ (q^2-k^2)^2 \sin (kd)\cosh (2qa),
\end{eqnarray*}
while $h_1 \equiv \alpha_0 \gamma_1 - \beta_0 \gamma_2 $ \,and\, $h_2 \equiv \alpha_0 \gamma_3 - \beta_0 \gamma_4$,
with
\begin{widetext}
\begin{eqnarray}
 \nonumber \gamma_1 &\equiv& -2k (q^2+k^2) \sin (kd) - d (q^2+k^2)^2 \cos (kd) + 2 q (q^2-k^2) \cos (kd) \sinh (2qa) \\
&-& 2qkd (q^2-k^2) \sin (kd) \sinh (2qa) - 2k (q^2-k^2)\sin (kd) \cosh (2qa) + d (q^2 -k^2)^2 \cos (kd) \cosh (2qa) ;\\ \nonumber \\
\nonumber  \gamma_2 &\equiv& 2kq \left\{ 2q\cos (kd) \cosh (2qa) -2qkd \sin (kd) \cosh (2qa) \right.\\
&-& \left. 2k \sin (kd) \sinh(2qa) + d (q^2-k^2)   \cos (kd) \sinh (2qa) \right\} \\ \nonumber \\
\nonumber  \gamma_3 &\equiv& -4q (q^2+k^2) \sin (kd) + 2k (3q^2-k^2) \cos (kd)\sinh (2qa) + 4kqa (q^2-k^2) \cos (kd) \cosh (2qa) ; \\
&+& 4q (q^2-k^2) \sin (kd) \cosh (2qa) + 2a (q^2-k^2)^2 \sin (kd) \sinh (2qa)  \\ \nonumber \\
\nonumber   \gamma_4 &\equiv&  2k \left\{ 4kq \cos (kd) \cosh (2qa) + (3q^2-k^2) \sin (kd) \sinh(2qa) \right.\\
  &+& \left. 4kq^2 a \cos (kd) \sinh (2qa) + 2qa (q^2-k^2) \sin (kd) \cosh(2qa) \right\}.
\end{eqnarray}
\end{widetext}

\end{document}